\documentclass[10pt]{article}
\usepackage{geometry}
\usepackage{xcolor}
\definecolor{graycolor}{gray}{0.9} 
\usepackage{microtype}
\usepackage{setspace} 
\usepackage{cmap}
\usepackage[utf8]{inputenc}
\usepackage[english]{babel}
\usepackage{times}
\usepackage{array}
\usepackage{soul}
\usepackage{cite}
\usepackage[numbers,sort&compress]{natbib}

\usepackage{titlesec} 
\titleformat {\section} [block] {\raggedright \fontsize{10}{10}\selectfont\bfseries} {\thesection. \space} {0pt} {}
\titlespacing {\section} {0pt} {12pt} {6pt}
\titleformat {\subsection} [block] {\raggedright \fontsize{10}{10}\selectfont\itshape} {\thesubsection .\space} {0pt} {}
\titlespacing {\subsection} {0pt} {12pt} {6pt}
\titleformat {\subsubsection} [block] {\raggedright \fontsize{10}{10}\selectfont} {\thesubsubsection .\space} {0pt} {}
\titlespacing {\subsubsection} {0pt} {12pt} {6pt}
\titleformat {\paragraph} [block] {\raggedright \fontsize{10}{10}\selectfont} {} {0pt} {}
\titlespacing {\paragraph} {0pt} {12pt} {6pt}

\usepackage{array} \newcommand{\PreserveBackslash}[1]{\let\temp=\\#1\let\\=\temp}
\newcolumntype{C}[1]{>{\PreserveBackslash\centering}m{#1}}
\newcolumntype{R}[1]{>{\PreserveBackslash\raggedleft}m{#1}}
\newcolumntype{L}[1]{>{\PreserveBackslash\raggedright}m{#1}}
\usepackage{lineno}
\usepackage{tabularx}
\usepackage{colortbl}
\usepackage{graphicx}
\usepackage{float}
\usepackage[export]{adjustbox}
\usepackage{caption}
\usepackage{fancyhdr} 
\usepackage{amsmath}
\usepackage{lastpage}
\usepackage{layout}
\usepackage{setspace} 
\usepackage{enumitem}
\usepackage{booktabs}
\usepackage{arydshln}
\usepackage{multirow}
\usepackage{color}
\usepackage{hyperref} 
\hypersetup{
	colorlinks=true,
	linkcolor=blue,
	filecolor=blue,
	urlcolor=black,
	citecolor=cyan,
}

\newcommand{\pacs}[1]{%
  \par\noindent\textbf{PACS:} #1\par
}

\setcitestyle{open={[},close={]},citesep={,\!},numbers}

\fancypagestyle{firstpage}{
    \setlength{\headsep}{2.2cm}
    
    \setlength{\footskip}{1.5cm}
    \fancyhf{}
    \lhead{\begin{table}[H]
        \centering
        \begin{tabular}{L{2.5cm}C{10cm}C{3.1cm}R{2cm}}
            \includegraphics[scale=0.035]{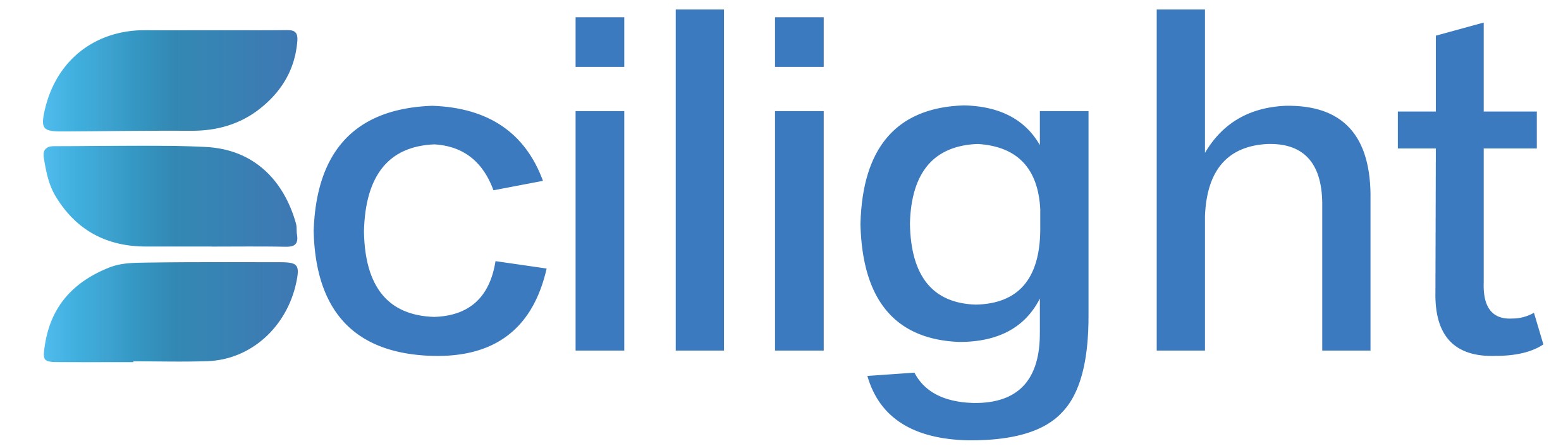} \vspace{-6pt}& \cellcolor{graycolor}\begin{tabular}[c]{@{}c@{}}\textit{International Journal of Gravitation and Theoretical Physics}\\ \href{https://www.sciltp.com/journals/ijgtp}{https://www.sciltp.com/journals/ijgtp}\end{tabular} & \includegraphics[scale=0.020]{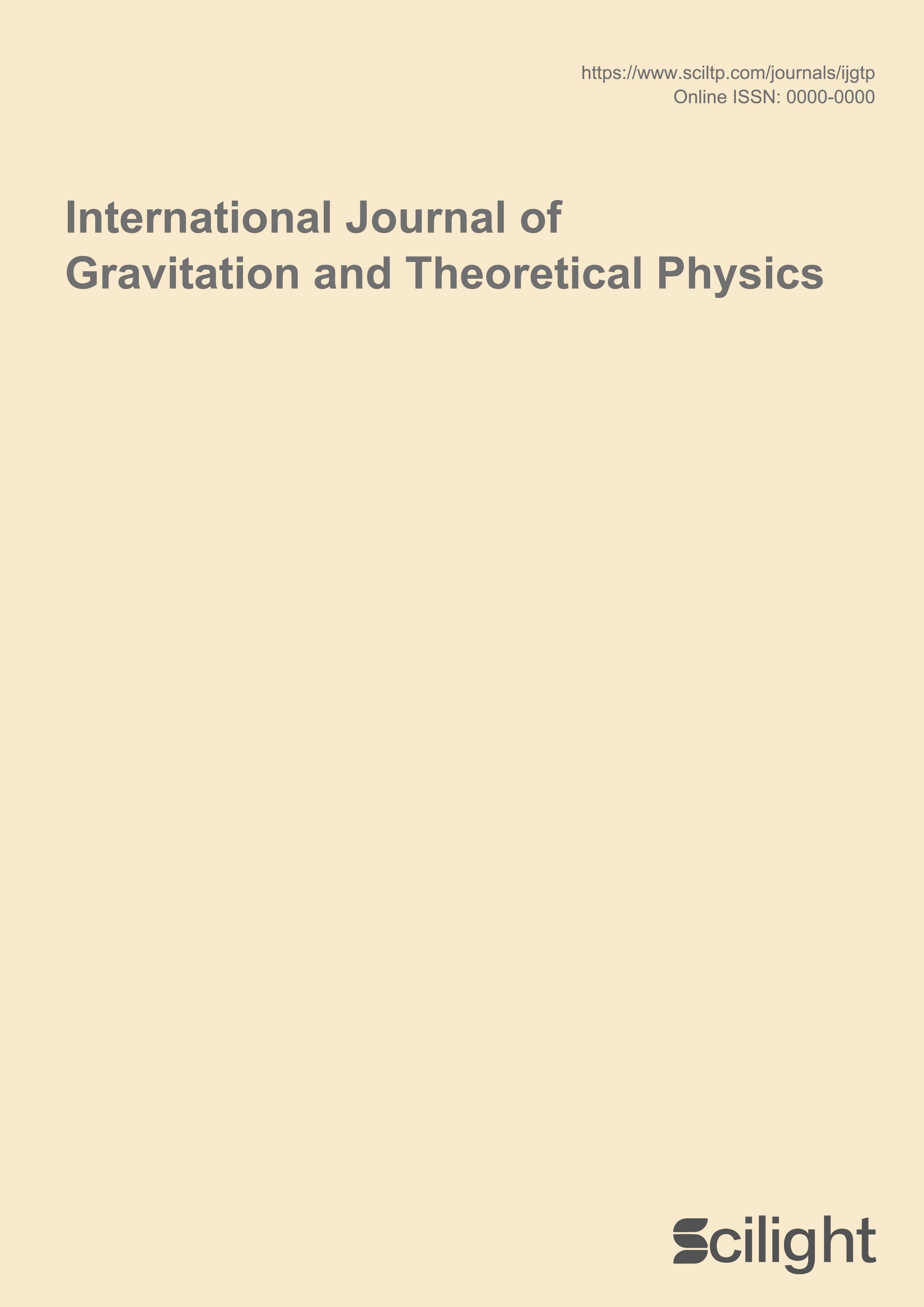} \vspace{-3pt}\\
        \end{tabular}
        \vspace{-22pt}
    \end{table}}
   
    \fancyfoot[C]{
        \vspace{-1.55cm}
        \begin{table}[H]
            \begin{minipage}[c]{0.15\columnwidth}
                \includegraphics[scale=0.5]{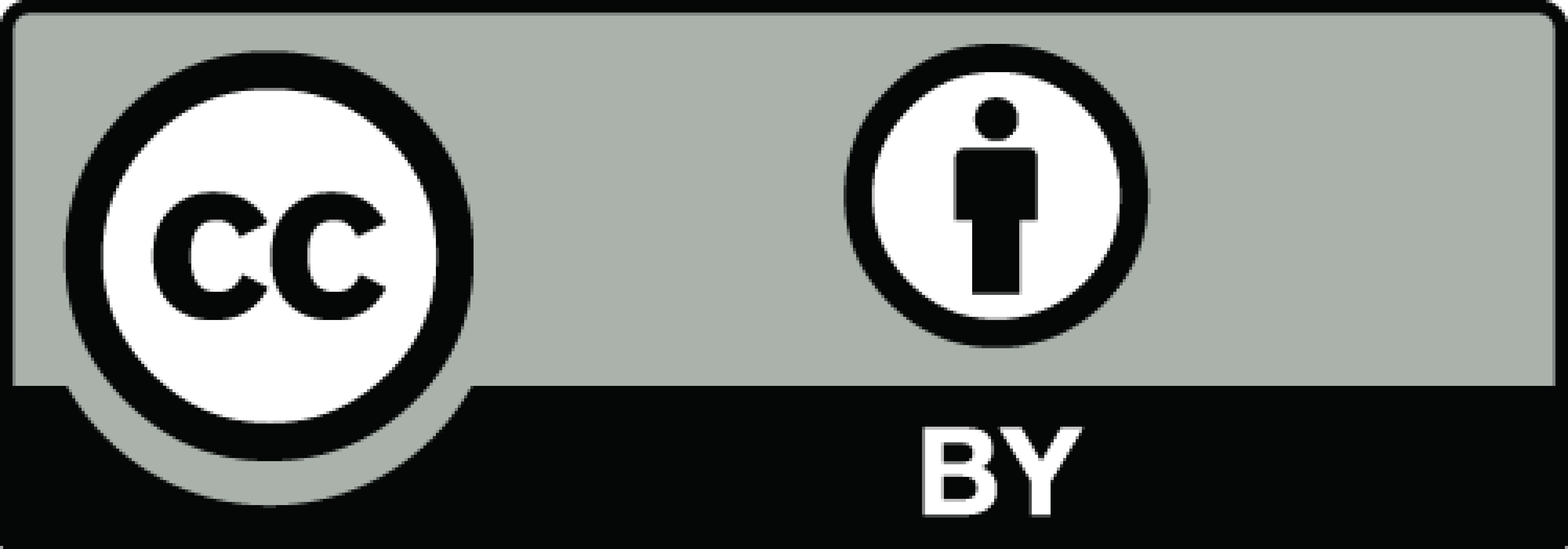} \vspace{1.1pt}
            \end{minipage}
            \hfill
            \begin{minipage}[c]{0.85\columnwidth}
                \scriptsize \textbf{Copyright:} © 2025 by the authors. This is an open access article under the terms and conditions of the Creative Commons Attribution (\mbox{CC BY}) license (\href{https://creativecommons.org/licenses/by/4.0/}{https://creativecommons.org/licenses/by/4.0/}). \\ \textbf{Publisher’s Note:} Scilight stays neutral with regard to jurisdictional claims in published maps and institutional affiliations.
            \end{minipage}
    \end{table}}
    \vspace{-0.55cm}
}

\begin{document}
\newgeometry{left=2.5cm, right=2.5cm, top=1.8cm, bottom=4cm}
	\thispagestyle{firstpage}
	\nolinenumbers
	{\noindent \textit{Article} }
	\vspace{4pt} \\
	{\fontsize{18pt}{10pt}\textbf{Grey-Body Factors for Scalar and Dirac Fields in the\\ \mbox{Euler-Heisenberg} Electrodynamics}  }
	\vspace{16pt} \\
	{\large Zainab Malik }
	\vspace{6pt}
	 \begin{spacing}{0.9}
		{\noindent \small
				Institute of Applied Sciences and Intelligent Systems, H-15 Islamabad, Pakistan; zainabmalik8115@outlook.com \vspace{6pt}\\
		\footnotesize	\textbf{How To Cite}: Malik, Z.  Grey-body factors for scalar and Dirac fields in the Euler-Heisenberg electrodynamics. \emph{International Journal of Gravitation and Theoretical Physics} \textbf{2025}, \emph{1}(1), 6. \href{https://doi.org/10.53941/ijgtp.2025.100006}{https://doi.org/10.53941/ijgtp.2025.100006}}.\\
	\end{spacing}

\begin{table}[H]
\noindent\rule[0.15\baselineskip]{\textwidth}{0.5pt} 
\begin{tabular}{lp{12cm}}  
 \small 
  \begin{tabular}[t]{@{}l@{}} 
  \footnotesize  Received: 6 August 2025 \\
  \footnotesize  Revised:  14 August 2025 \\
   \footnotesize Accepted: 27 August 2025 \\
  \footnotesize  Published: 16 September 2025
  \end{tabular} &
  \textbf{Abstract:} We study grey-body factors of neutral scalar and Dirac fields in the background of charged black holes arising in the Einstein--Euler--Heisenberg (EEH) theory. The Euler--Heisenberg corrections, which represent nonlinear electrodynamical effects due to vacuum polarization, modify the effective potential barrier surrounding the black hole and thereby affect the transmission probabilities 
for Hawking quanta. Using the sixth-order WKB method, and verifying our results against the recently proposed correspondence between grey-body factors and quasinormal modes, we compute the frequency-dependent grey-body spectra for various values of the black hole charge and EEH coupling. We find that the nonlinear coupling systematically lowers the effective potential barrier, enhancing the transmission probability. This work represents the first step toward incorporating nonlinear QED effects into the study of Hawking evaporation, focusing here on neutral test fields.  \\
\\
  & 
   \vspace{6pt}
 \\

  &  \pacs{04.30.-w; 04.50.Kd; 04.70.-s}
\end{tabular}
\noindent\rule[0.15\baselineskip]{\textwidth}{0.5pt} 
\end{table}

	\section{Introduction }
	The study of grey-body factors of black holes has become an essential part of modern black hole physics, as they encode the frequency-dependent transmission probabilities of radiation through the effective potential barrier surrounding the black hole. While Hawking radiation \cite{Hawking:1975vcx} is often approximated as purely thermal, the actual spectrum is modified by these grey-body factors, which depend sensitively on the black hole geometry, the nature of the perturbing field, and the underlying gravitational and matter theory \cite{Page:1976df,Page:1976ki,Kanti:2004nr}. Thus, understanding grey-body factors is indispensable both for probing quantum properties of black holes and for identifying possible signatures of physics beyond the standard Einstein–Maxwell framework.

Despite the existence of numerous exotic models of nonlinear electrodynamics, many of which either lack a well-defined weak-field limit or are introduced purely as phenomenological extensions, in this work we restrict our attention to the physically well-motivated Euler--Heisenberg electrodynamics \cite{Heisenberg:1936nmg}. One of the most remarkable predictions of the Euler--Heisenberg Lagrangian is the possibility of light-by-light (photon--photon) scattering in vacuum, a purely quantum effect absent in classical electrodynamics. The theoretical foundations for this process were first laid in \cite{Karplus:1950zz} and have since been developed extensively, with modern collider experiments providing growing evidence for its realization \cite{dEnterria:2013zqi,TOTEM:2021zxa}. Consequently, Euler--Heisenberg electrodynamics serves not only as a theoretically robust model but also as a phenomenologically relevant one, firmly rooted in the established framework of QED. For these reasons, it constitutes a natural and compelling setting in which to explore the grey-body factors and Hawking radiation of charged black holes. When coupled to Einstein gravity, the resulting Einstein–Euler–Heisenberg (EEH) theory describes charged black holes whose structure departs from the standard Reissner–Nordström solutions. These deviations, though typically small, may have profound implications for both the stability and observable properties of such black holes \cite{Bolokhov:2024ixe,Nomura:2021efi,Breton:2021mju,Breton:2016mqh}.

\textls[-15]{From the perspective of Hawking radiation, the presence of nonlinear electromagnetic corrections modifies the effective potential felt by quantum fields, and consequently, the grey-body factors. This raises several important~questions:}
\restoregeometry

\noindent How do QED-induced nonlinearities alter the transmission probabilities compared to the classical  Einstein–Maxwell case? Can these modifications be significant in the strong-field regime near the event horizon? 

In addition, grey-body factors in the EEH background are closely connected to the quasinormal mode spectrum, via the relation between transmission/reflection amplitudes and the poles of the scattering matrix \cite{Magos:2020ykt}. Thus, the study of grey-body factors not only complements the analysis of black hole perturbations but also provides an independent check of the stability and dynamical response of EEH black holes. Finally, as grey-body factors enter directly into the calculation of energy and charge emission rates, their study is essential for realistic modeling of Hawking evaporation in nonlinear electrodynamics settings.
In this work, we make the first step in this direction and focus on the grey-body factors of neutral scalar and Dirac field in the background of the charged black holes in the Einstein–Euler–Heisenberg theory.

\section{The Black Hole Background}

Black hole spacetimes arising from nonlinear electrodynamics (NED) have received considerable attention as physically motivated modifications of classical general relativity. Unlike Maxwell’s theory, certain NED models are capable of producing regular or modified black hole solutions without invoking exotic matter or quantum gravity corrections. One prominent example is the Euler–Heisenberg effective Lagrangian, which encapsulates quantum corrections to electrodynamics due to vacuum polarization in strong fields and leads to modified electromagnetic dynamics at the classical level.

In this work, we study the Euler–Heisenberg electrodynamics coupled to general relativity. This setting yields a static, spherically symmetric black hole solution that modifies the classical Reissner–Nordström geometry at small radii while preserving asymptotic flatness. The corresponding action takes the form \cite{Magos:2020ykt}
\begin{equation}
S = \frac{1}{4\pi} \int_{M^{4}} d^{4}x \sqrt{-g} 
\left[ \frac{1}{4} R   +F - \frac{4\alpha^2}{45m_e^4}\,F^{2} - \frac{7\alpha^2}{45m_e^4}\,G^{2} \right],
\end{equation}
where $g$ is the determinant of the metric tensor, $R$ is the Ricci scalar, and $F \equiv \tfrac{1}{4}F_{\mu\nu}F^{\mu\nu}$, $G \equiv \tfrac{1}{4}F_{\mu\nu}\,{}^{\ast}F^{\mu\nu}$, with $F_{\mu\nu}$ standing for the electromagnetic field strength tensor; $\alpha$ is
the fine structure constant and $m_e$ is the electron~mass.

The static, spherically symmetric solution sourced by a purely electric field in this theory reads:
\begin{equation}
ds^2 = -f(r)\, dt^2 + \frac{dr^2}{f(r)} + r^2 d\Omega^2,
\end{equation}
where the lapse function is given by
\begin{equation}
f(r) = 1 - \frac{2M}{r} + \frac{Q^2}{r^2} - \frac{a Q^4}{r^6},
\label{metric-function}
\end{equation}
where $a$ is the constant related to the coupling $\alpha$ as follows:
\begin{equation}
    a\equiv\frac{32\alpha^2}{9m_e^4}.
\end{equation}

Here, $M$ and $Q$ represent the ADM mass and electric charge of the black hole, respectively. The final term arises from the Euler–Heisenberg correction and becomes significant in the strong-field regime near the black hole.

In the limit $a \to 0$, the standard Reissner–Nordström solution is recovered. However, for finite $a$, the effective geometry exhibits important deviations in the near-horizon region, modifying both the location of the horizons and the behavior of perturbations. Importantly, the solution remains asymptotically flat, making it a suitable setting for studying quasinormal ringing using standard boundary conditions \cite{Bolokhov:2024ixe}.

To probe the dynamics of this background, we consider linear test field perturbations of scalar and Dirac types. The wave equations for these fields reduce, after suitable separation of variables, to Schrödinger-like forms:
\begin{equation}
\frac{d^2 \Psi}{dr_*^2} + \left[\omega^2 - V\right]\Psi = 0,
\end{equation}
where $r_*$ is the tortoise coordinate defined by $dr_*/dr = 1/f(r)$, and $V(r)$ is the effective potential depending on the spin of the field and the background geometry.

For a minimally coupled scalar field, the effective potential is given by
\begin{equation}
V_\text{s}(r) = f(r)\left[\frac{j(j+1)}{r^2} + \frac{f'(r)}{r} \right],
\end{equation}
where $j$ is the multipole number and $\mu$ is the scalar field mass.

For a massive Dirac field propagating in the background metric \eqref{metric-function},  the effective potentials $V_\pm(r)$ are given by
\begin{equation}
V_\pm(r) = W(r)^2 \pm \frac{dW(r)}{dr_*}.
\end{equation}

The superpotential $W(r)$ for a massive Dirac field takes the form
\begin{equation}
W(r) =\frac{\kappa}{r}  \sqrt{f(r)} ,
\end{equation}
where $\kappa = \pm(j + 1/2)$ with $j=1/2,3/2,5/2,...$ being the total angular momentum quantum number, and $\mu$ is the mass of the Dirac field. The two potentials $V_+(r)$ and $V_-(r)$ are supersymmetric-like partners and lead to the same spectrum of quasinormal frequencies.

In the following sections, we aim to calculate grey-body factors of the scalar and Dirac fields in the background of the charged black hole with corrections originating from the non-linear electrodynamics.

\section{Grey-Body Factors and the WKB Method}

Hawking radiation emitted by black holes is not exactly thermal. 
The radiation spectrum measured at infinity is filtered by the effective potential 
barrier surrounding the black hole, so that only a portion of the originally produced 
particles escape to infinity. This modification of the pure thermal spectrum is described 
by the so-called grey-body factors. For a given frequency $\omega$ and multipole number $j$, 
the grey-body factor $\Gamma_{j}(\omega)$ is defined as the transmission probability 
for a wave to tunnel through the effective potential barrier. Thus, the differential energy 
emission rate of a black hole takes the form \cite{Hawking:1975vcx}
\begin{equation}
\frac{d^{2}E}{d\omega\,dt} = 
\sum_{j} \frac{\Gamma_{j}(\omega)}{2\pi}
\frac{\omega}{e^{\omega/T_{H}} \mp 1},
\end{equation}
where $T_{H}$ is the Hawking temperature and the sign in the denominator corresponds 
to bosonic ($-$) or fermionic ($+$) statistics, respectively, for the scalar and Dirac field. The grey-body factors therefore play 
a crucial role in determining both the intensity and spectral shape of Hawking radiation.

The asymptotic boundary conditions are
\begin{equation}
\Psi(r_{*}\to -\infty) \sim e^{-i\omega r_{*}}, 
\qquad 
\Psi(r_{*}\to +\infty) \sim A_{\text{out}}\,e^{+i\omega r_{*}}
+ A_{\text{in}}\,e^{-i\omega r_{*}}.
\end{equation}

The grey-body factor is then given by
\begin{equation}
\Gamma_{j}(\omega) = 
\left|\frac{A_{\text{trans}}}{A_{\text{in}}}\right|^{2}
= 1 - \left|\frac{A_{\text{out}}}{A_{\text{in}}}\right|^{2},
\end{equation}
where $A_{\text{in}}$ and $A_{\text{out}}$ are the amplitudes of the incoming 
and reflected waves, respectively, and $A_{\text{trans}}$ corresponds to the transmitted flux.

Direct numerical integration of the wave equation provides precise values for 
$\Gamma_{j}(\omega)$, but an efficient and widely used semi-analytic method 
is the Wentzel–Kramers–Brillouin (WKB) approximation. The WKB technique, 
originally adapted to calculations of quasinormal modes and grey-body factors of black holes in~\cite{Schutz:1985km,Iyer:1986np,Konoplya:2003ii}, 
yields analytic expressions for the transmission and reflection coefficients by 
matching the asymptotic solutions across the classically forbidden region 
around the potential barrier.

In its commonly employed form, the WKB method expands the transmission coefficient 
near the peak of the effective potential. At leading order, the grey-body factor is given by \cite{Schutz:1985km}
\begin{equation}
\Gamma_{j}(\omega) \approx 
\frac{1}{1 + \exp\!\left(2\pi i K\right)},
\end{equation}
where, in the eikonal limit $K$ depends on $V_{0}$ the height of the potential barrier and $V_{0}''$ the second derivative of the potential with respect to the tortoise coordinate evaluated at the maximum. Higher-order corrections, 
incorporating third and higher derivatives of $V(r)$, systematically improve the accuracy 
of the method and are usually expressed via Padé resummation~\cite{Matyjasek:2017psv,Konoplya:2019hlu}.
Here we used the 6th order formula \cite{Konoplya:2003ii} without using Padé resummation, because it is not straighforward for grey-body factors, unlike the case of quasinormal modes \cite{Matyjasek:2017psv}. This method has been widely used for finding quasinormal modes and grey-body factors in numerous publications (see, for recent examples \cite{Matyjasek:2021xfg,Malik:2023bxc,Bolokhov:2023dxq,Skvortsova:2023zmj,Malik:2024nhy} and a review \cite{Konoplya:2019hlu}).

The WKB approach provides reliable results for moderate to large multipole numbers $j$ 
and for frequencies near or above the peak of the effective potential, 
making it particularly suitable for estimating grey-body factors in 
the semi-classical regime relevant to Hawking evaporation. 
Although less accurate for very low frequencies, where numerical methods are preferable, 
it remains a valuable analytic tool that offers physical insight into the scattering process. The analytic approximation for the quasinormal modes of a scalar field can be seen in Appendix A.

Recent work~\cite{Konoplya:2024lir} has established a direct correspondence between grey-body factors and the lowest quasinormal modes of black holes. In the eikonal regime, the grey-body factor can be expressed as
\[
\Gamma_{j}(\Omega) \approx \left[ 1 + \exp\!\left( \frac{2\pi\bigl(\Omega^{2}-\mathrm{Re}(\omega_{0})^{2}\bigr)}{4\,\mathrm{Re}(\omega_{0})\,\mathrm{Im}(\omega_{0})} \right) \right]^{-1},
\]
where $\omega_{0}$ is the fundamental quasinormal frequency. For moderate values of the multipole number $j$, the accuracy of this formula can be significantly improved by including the first overtone $\omega_{1}$ via the correction
\[
-iK = -\frac{\Omega^{2}-\mathrm{Re}(\omega_{0})^{2}}{4\,\mathrm{Re}(\omega_{0})\,\mathrm{Im}(\omega_{0})} + \Delta_{1}, 
\qquad 
\Delta_{1} = \frac{\mathrm{Re}(\omega_{0})-\mathrm{Re}(\omega_{1})}{16\,\mathrm{Im}(\omega_{0})},
\]
so that $\Gamma_{j}(\Omega) = [1+e^{2\pi iK}]^{-1}$. Higher-order refinements involving additional terms $\Delta_{2}$ and $\Delta_{f}$ further enhance the precision, showing that the grey-body spectrum is largely determined by the fundamental mode and its first overtone. This correspondence has been tested for a number of cases \cite{Malik:2024cgb,Skvortsova:2024msa,Dubinsky:2024vbn,Lutfuoglu:2025ldc,Lutfuoglu:2025ohb,Lutfuoglu:2025ljm} and extended to rotating black holes \cite{Konoplya:2024vuj} and wormholes \cite{Bolokhov:2024otn}.

\section{Interpretation of Results}

The grey-body factors obtained in the Einstein--Euler--Heisenberg black hole background
exhibit a consistent dependence on the physical parameters of the system, and their behavior
can be directly traced to modifications of the corresponding effective potentials.

\begin{figure}[H]
\centering
\resizebox{0.72 \linewidth}{!}{\includegraphics{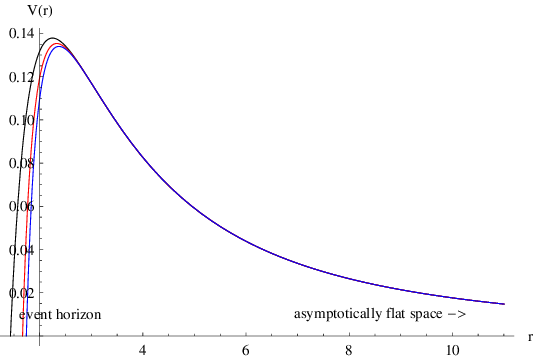}}
\caption{Effective potentials for $j=1$ scalar perturbations: $M=1$, $Q=0.902$, $a=0$ (black), $a=3$ (red), and $a=5$ (blue).}\label{fig:Potentials}
\end{figure}

\begin{figure}[H]
\centering
\resizebox{0.72 \linewidth}{!}{\includegraphics{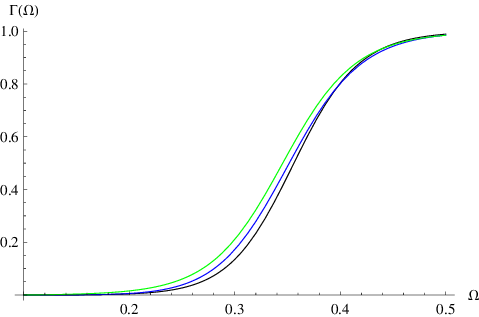}}
\caption{Grey-body factors for $j=0$ scalar perturbations calculated by the 6th order WKB formula. Here we have $M=1$, $Q=0.902$, $a=0$ (black), $a=1$ (blue), and $a=3$ (green).}\label{fig:L0variousA}
\end{figure}

Figure~\ref{fig:Potentials} illustrates the effect of the Euler--Heisenberg coupling $a$
on the effective potential for scalar perturbations with $j=1$. As $a$ increases, the
height of the potential barrier decreases, especially in the near-horizon region,
leading to enhanced transmission probabilities. This trend is reflected in the
grey-body factors for $j=0$ scalar perturbations shown in Figure~\ref{fig:L0variousA},
where larger values of $a$ correspond to systematically larger grey-body factors across
the frequency range considered.

\begin{figure}[H]
\centering
\resizebox{0.95\linewidth}{!}{\includegraphics{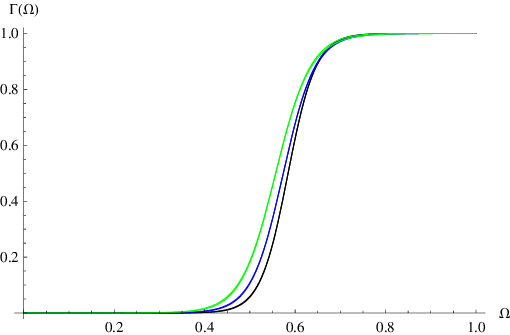}\includegraphics{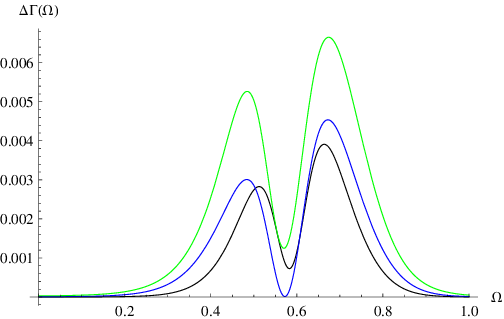}}
\caption{Grey-body factors for $j=2$ scalar perturbations calculated by the 6th order WKB formula and by the correspondence with QNMs (\textbf{left}), and the absolute difference between the data obtained by the two methods (\textbf{right}). Here we have $M=1$, $a=0$ (black), $a=3$ (blue), and $a=10$ (green).}\label{fig:L2variousA}
\end{figure}

\begin{figure}[H]
\centering
\resizebox{0.95\linewidth}{!}{\includegraphics{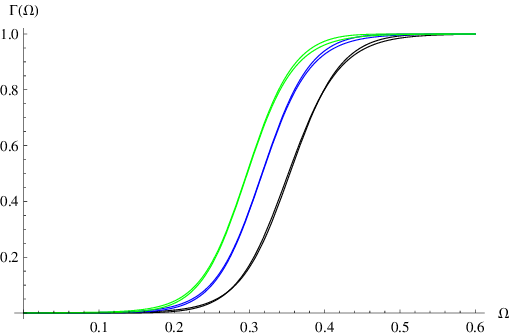}\includegraphics{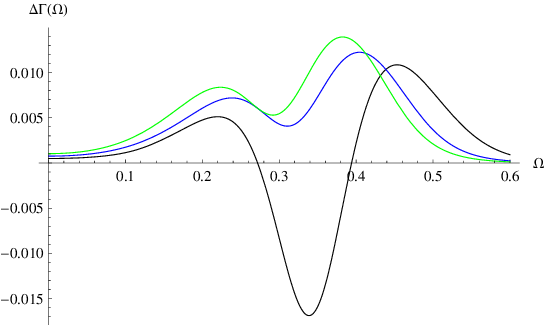}}
\caption{Grey-body factors for $j=1$ scalar perturbations calculated by the 6th order WKB formula and by the correspondence with QNMs (\textbf{left}), and the absolute difference between the data obtained by the two methods (\textbf{right}). Here we have $M=1$, $a=1$, $Q=0.902$ (black), $Q=0.6$ (blue), and $Q=0$  (green).}\label{fig:L1variousQ}
\end{figure}

\begin{figure}[H]
\centering
\resizebox{0.95\linewidth}{!}{\includegraphics{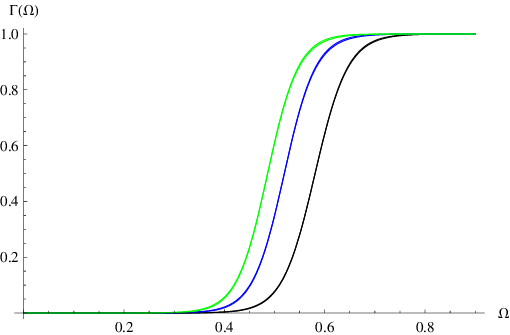}\includegraphics{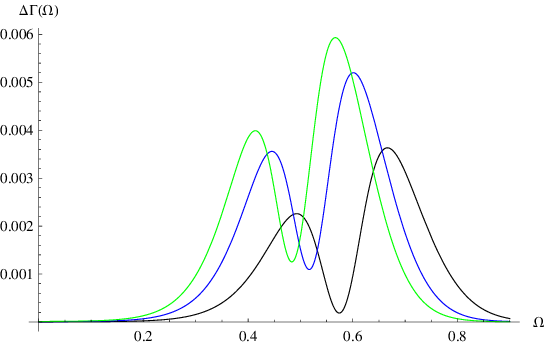}}
\caption{Grey-body factors for $j=2$ scalar perturbations calculated by the 6th order WKB formula and by the correspondence with QNMs (\textbf{left}), and the absolute difference between the data obtained by the two methods (\textbf{right}). Here we have $M=1$, $a=1$, $Q=0.902$ (black), $Q=0.6$ (blue), and $Q=0$  (green).}\label{fig:L2variousQ}
\end{figure}

The comparison between the sixth-order WKB approximation and the correspondence with
quasinormal modes (QNM) is shown in Figures~\ref{fig:L2variousA}--\ref{fig:L2variousQ}.
For both $j=1$ and $j=2$ scalar perturbations, the agreement between the two methods is
remarkably good in the frequency domain where the WKB approximation is expected to be
reliable. The right panels demonstrate that the discrepancies remain small, typically
at the sub-percent level, confirming the robustness of the QNM correspondence in this
nonlinear electrodynamics setting. The effect of the black hole charge $Q$ is also
evident: increasing $Q$ raises the potential barrier, thereby reducing transmission,
but this suppression can be mitigated by higher values of the Euler--Heisenberg coupling.

\begin{figure}[H]
\centering
\resizebox{0.95\linewidth}{!}{\includegraphics{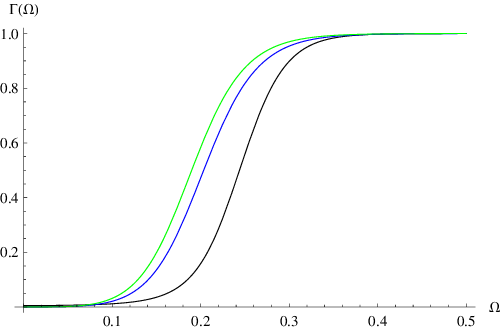}\includegraphics{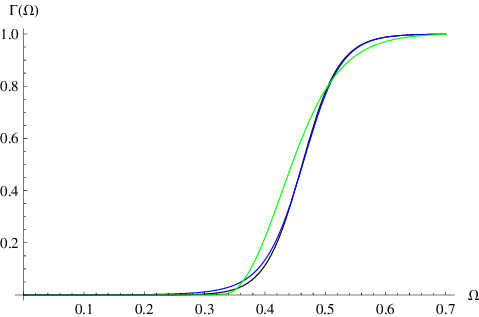}}
\caption{(\textbf{Left}): Grey-body factors for $j=1/2$ Dirac perturbations calculated by the 6th order WKB formula. Here we have $M=1$, $a=1$, $Q=0.902$ (black), $Q=0.6$ (blue), and $Q=0$  (green). (\textbf{Right}): Grey-body factors for $j=3/2$ Dirac perturbations calculated by the 6th order WKB formula. Here we have $M=1$, $Q=0.902$, $a=0$ (black), $a=3$ (blue), and $a=8$  (green).}\label{fig:DiracVariousA}
\end{figure}

\begin{figure}[H]
\centering
\resizebox{0.95\linewidth}{!}{\includegraphics{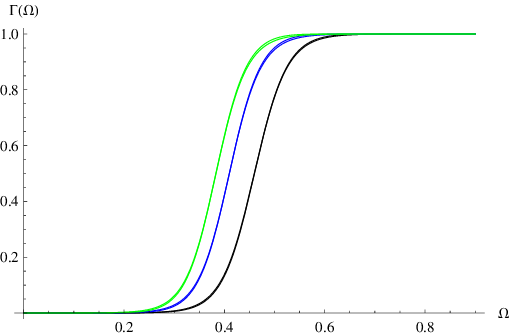}\includegraphics{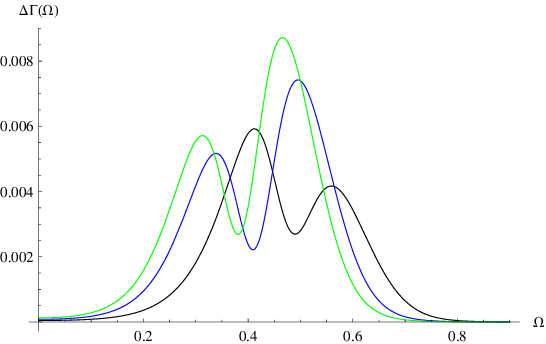}}
\caption{Grey-body factors for $j=3/2$ Dirac perturbations calculated by the 6th order WKB formula and by the correspondence with QNMs (\textbf{left}), and the absolute difference between the data obtained by the two methods (\textbf{right}). Here we have $M=1$, $a=1$, $Q=0.902$ (black), $Q=0.6$ (blue), and $Q=0$  (green).}\label{fig:L32variousQ}
\end{figure}

For Dirac perturbations, the qualitative picture is similar. As shown in
Figure~\ref{fig:DiracVariousA}, both $j=1/2$ and $j=3/2$ modes exhibit larger
grey-body factors when the coupling $a$ is increased, reflecting the lowering of
the effective barrier. The role of the black hole charge is demonstrated in
Figure~\ref{fig:L32variousQ}, where decreasing $Q$ enhances the transmission, while
the comparison with the QNM correspondence again shows excellent agreement within
the WKB validity range. The supersymmetric nature of the Dirac potentials ensures
that $V_{+}(r)$ and $V_{-}(r)$ lead to identical grey-body factors, which we have
confirmed numerically.

In summary, the results show that nonlinear QED corrections encoded in the
Euler--Heisenberg coupling $a$ systematically enhance grey-body factors for both
scalar and fermionic fields. Since grey-body factors directly determine the intensity
and spectral distribution of Hawking radiation, this implies that charged black holes
in the Einstein--Euler--Heisenberg theory may evaporate more efficiently than their
Reissner--Nordström counterparts. The good agreement with the QNM correspondence
further supports the reliability of semi-analytic approaches in this~context.

\vspace{-24pt}

\vspace{-24pt}

\vspace{-24pt}

\vspace{-24pt}

\vspace{-24pt}

\vspace{-24pt}

\vspace{-24pt}

\section{Conclusions}

In this work we have carried out the first analysis of grey-body factors for 
neutral scalar and Dirac fields in the background of charged black holes within 
the Einstein--Euler--Heisenberg theory. By employing the sixth-order WKB 
approximation and comparing with the recently established correspondence between 
grey-body factors and the lowest quasinormal modes, we have demonstrated that 
nonlinear QED corrections, parameterized by the coupling $a$, systematically 
enhance the transmission probabilities through the effective potential barrier. 
This enhancement is especially significant in the low-frequency regime, where 
the Hawking emission spectrum is most sensitive to grey-body suppression. 

Our study shows that the grey-body factors in the EEH background are largely 
determined by the interplay between the black hole charge $Q$ and the Euler--Heisenberg 
coupling $a$, with the latter reducing the effective barrier height and thereby 
amplifying the emission rate. Since grey-body factors enter directly into the 
calculation of Hawking fluxes, the results obtained here provide a useful basis 
for estimating the intensity and spectral distribution of Hawking radiation in 
nonlinear electrodynamics scenarios.

It should be emphasized that the present analysis focuses only on neutral test 
fields. A more realistic and physically rich problem is the study of \emph{charged} 
perturbations in the Einstein--Euler--Heisenberg background. The inclusion of 
field--background interactions, charge coupling, and superradiant effects is 
expected to lead to qualitatively new phenomena. We leave this direction for 
future work, noting that such an extension would be essential for a complete 
understanding of evaporation processes and possible observational signatures of 
nonlinear electrodynamics around black holes.

Thus, our results represent an initial step towards incorporating quantum 
electrodynamical corrections into black hole radiation spectra. By combining 
grey-body factors with the established framework for quasinormal modes, 
this approach paves the way for future studies aimed at refining the 
predictions of Hawking evaporation rates in strongly curved and 
nonlinearly corrected spacetimes. 
	

		\section*{Funding}
This research received no external funding. 
 
		\section*{Institutional Review Board Statement}
Not applicable.

		\section*{Informed Consent Statement}
Not applicable.

		\section*{Data Availability Statement}
Not applicable.

		\section*{Conflicts of Interest}
The author declares no conflict of interest.
	
	\section*{Appendix A. Analytic Expression for the Scalar Quasinormal Modes}
	
	Using the general approach developed in \cite{Konoplya:2023moy} we expand the position of the potential peak into the powers of $\kappa^{-1}=(j+1/2)^{-1}$ and $Q$,

\begin{equation}\label{rmax-scalar-mu0}
\begin{array}{rcl}\nonumber
r_{\max } &=& \displaystyle -\frac{M}{3 \kappa ^2}+3 M
+Q^2 \left(\frac{5}{27 M\kappa ^2}-\frac{2}{3 M}\right)\\
&&\displaystyle
+Q^4 \left(\frac{4 \left(a-9 M^2\right)}{243 M^5}
+\frac{-\tfrac{13 a}{M^2}+36}{729 M^3 \kappa ^2}\right)
+\mathcal{O}\!\left(Q^6,\frac{1}{\kappa ^4}\right).
\end{array}
\end{equation}
Then, using the higher order WKB formula for the quasinormal frequency $\omega$, we find,
\begin{small}
\begin{equation}\label{eikonal-scalar-mu0}
\begin{array}{rcl}\nonumber
\omega  &=& \displaystyle
\frac{i K \left(-940 K^2-313\right)}{46656 \sqrt{3} M \kappa ^2}
+\frac{-60 K^2+29}{1296 \sqrt{3} M \kappa }
+\frac{\kappa }{3 \sqrt{3} M}
-\frac{i K}{3 \sqrt{3} M}\\
&&\displaystyle+Q^2 \left(
\frac{\kappa }{18 \sqrt{3} M^3}
-\frac{i K}{54 \sqrt{3} M^3}
+\frac{i K \left(1100 K^2+305\right)}{279936 \sqrt{3} M^3 \kappa ^2}
+\frac{204 K^2+31}{23328 \sqrt{3} M^3 \kappa }
\right)\\
&&\displaystyle+ Q^4 \left(
-\frac{i K \left(a \left(60400 K^2 +114196\right) 
+9 M^2\left(-9868 K^2 -3553\right)\right)}{30233088 \sqrt{3} M^7 \kappa ^2}\right.\\
&&\displaystyle\left.+\frac{4 a \left(300 K^2+1223\right) 
-9 M^2\left(-5244 K^2+317\right)}{7558272 \sqrt{3} M^7 \kappa } \right.\\
&&\displaystyle\left.
+ \frac{\kappa (117 M^2 -4 a)}{5832 \sqrt{3} M^7}
+ \frac{i K (81 M^2 -68 a)}{17496 \sqrt{3} M^7}
\right)
+ \mathcal{O}\!\left(Q^6,\frac{1}{\kappa ^3}\right),
\end{array}
\end{equation}
\end{small}
for the quasinormal modes (where $\kappa\equiv j+1/2$, $K\equiv n+1/2$).
Using the correspondence between quasinormal modes and grey-body factors, one can easily find an approximate analytic expression for the latter.
However, for the Dirac fields the analytic expressions are cumbersome.

	\small
	\bibliographystyle{scilight}

\end{document}